\RequirePackage{amsthm}
\documentclass[pdflatex,sn-nature,Numbered]{sn-jnl}

\usepackage{graphicx}%
\usepackage{multirow}%
\usepackage{makecell}
\usepackage{amsmath,amssymb,amsfonts}%
\usepackage{amsthm}%
\usepackage{mathrsfs}%
\usepackage[title]{appendix}%
\usepackage{xcolor}%
\usepackage{textcomp}%
\usepackage{manyfoot}%
\usepackage{booktabs}%
\usepackage{algorithm}%
\usepackage{algorithmicx}%
\usepackage{algpseudocode}%
\usepackage{listings}%
\usepackage{orcidlink}

\makeatletter
\newcommand{\currentfsize}{\f@size pt}
\makeatother

\newdimen\fsize
\newcommand{\setfsize}{\setlength{\fsize}{\currentfsize}}

\theoremstyle{thmstyleone}%
\theoremstyle{thmstyletwo}%

\theoremstyle{thmstylethree}%

\begin{document}

\title[The Simons Observatory: Large-Scale Characterization of 90/150 GHz TES Detector Modules]{The Simons Observatory: Large-Scale Characterization of 90/150 GHz TES Detector Modules}


\author*[1]{\fnm{Daniel} \sur{Dutcher} \orcidlink{0000-0002-9962-2058}}\email{ddutcher@princeton.edu}

\author[2]{\fnm{Shannon M.} \sur{Duff} \orcidlink{0000-0002-9693-4478}}

\author[2,3]{\fnm{John C.} \sur{Groh} \orcidlink{0000-0001-9880-3634}}

\author[4]{\fnm{Erin} \sur{Healy} \orcidlink{0000-0002-3757-4898}}

\author[2]{\fnm{Johannes} \sur{Hubmayr} \orcidlink{0000-0002-2781-9302}}

\author[5]{\fnm{Bradley R.} \sur{Johnson} \orcidlink{0000-0002-6898-8938}}

\author[2,3]{\fnm{Dante} \sur{Jones} \orcidlink{0009-0005-1195-2458}}

\author[6]{\fnm{Ben} \sur{Keller} \orcidlink{0000-0002-2978-7957}}

\author[6]{\fnm{Lawrence T.} \sur{Lin} \orcidlink{0000-0001-5465-8973}}

\author[2]{\fnm{Michael J.} \sur{Link} \orcidlink{0000-0003-2381-1378}}

\author[2]{\fnm{Tammy J.} \sur{Lucas} \orcidlink{0000-0001-7694-1999}}

\author[1]{\fnm{Samuel} \sur{Morgan}}

\author[1]{\fnm{Yudai} \sur{Seino} \orcidlink{0000-0001-5680-4989}}

\author[1]{\fnm{Rita F.} \sur{Sonka} \orcidlink{0000-0002-1187-9781}}

\author[1]{\fnm{Suzanne T.} \sur{Staggs} \orcidlink{0000-0002-7020-7301}}

\author[1]{\fnm{Yuhan} \sur{Wang} \orcidlink{0000-0002-8710-0914}}

\author[1]{\fnm{Kaiwen} \sur{Zheng} \orcidlink{0000-0003-4645-7084}}

\affil*[1]{\orgdiv{Department of Physics}, \orgname{Princeton University}, \orgaddress{\street{Jadwin Hall}, \city{Princeton}, \postcode{08540}, \state{NJ}, \country{USA}}}

\affil[2]{\orgdiv{Quantum Sensors Division}, \orgname{NIST}, \orgaddress{\street{325 Broadway}, \city{Boulder}, \postcode{80305}, \state{CO}, \country{USA}}}

\affil[3]{\orgdiv{Department of Physics}, \orgname{University of Colorado}, \orgaddress{\city{Boulder}, \postcode{80309}, \state{CO}, \country{USA}}}

\affil[4]{\orgdiv{Kavli Institute for Cosmological Physics}, \orgname{University of Chicago}, \orgaddress{\street{5640 South Ellis Avenue}, \city{Chicago}, \postcode{60637}, \state{IL}, \country{USA}}}

\affil[5]{\orgdiv{Department of Astronomy}, \orgname{University of Virginia}, \orgaddress{\city{Charlottesville}, \postcode{22904}, \state{VA}, \country{USA}}}

\affil[6]{\orgdiv{Department of Physics}, \orgname{Cornell University}, \orgaddress{\city{Ithaca}, \postcode{14853}, \state{NY}, \country{USA}}}

\abstract{The Simons Observatory (SO) is a cosmic microwave background instrumentation suite being deployed in the Atacama Desert in northern Chile.
The telescopes within SO use three types of dichroic transition-edge sensor (TES) detector arrays, with the 90 and 150~GHz Mid-Frequency (MF) arrays containing 65\% of the approximately 68,000 detectors in the first phase of SO.
All of the 26 required MF detector arrays have now been fabricated, packaged into detector modules, and tested in laboratory cryostats.
Across all modules, we find an average operable detector yield of 84\% and median saturation powers of (2.8, 8.0)~pW  with interquartile ranges of (1, 2)~pW at (90, 150)~GHz, respectively, falling within their targeted ranges.
We measure TES normal resistances and superconducting transition temperatures on each detector wafer to be uniform within 3\%, with overall central values of  7.5~m$\Omega$ and  165~mK, respectively.
Results on time constants, optical efficiency, and noise performance are also presented and are consistent with achieving instrument sensitivity forecasts.
}


\maketitle

\section{Introduction}\label{sec1}

The Simons Observatory (SO) is a suite of telescopes optimized to measure the cosmic microwave background, sited on Cerro Toco (elevation 5200~m) in the Atacama Desert in northern Chile \cite{thesimonsobservatorycollaboration2019b}.
SO is designed to observe the millimeter-wave sky over a wide range of angular scales and frequencies, with spectral coverage in six frequency bands distributed across three types of dichroic transition-edge sensor (TES) detector arrays: Low-Frequency (LF) arrays with approximate band centers near 30 and 40 GHz, Mid-Frequency (MF) arrays at 90 and 150 GHz, and Ultra-High-Frequency (UHF) arrays at 220 and 280 GHz. Each of the MF and UHF arrays contain 430 dichroic, dual-polarization pixels with 4 TESs per pixel, plus 36 non-optically-coupled calibration TESs, for a total of 1756 TESs per detector wafer.
The SO MF and UHF detector wafer fabrication process is discussed in \cite{duff2023}.

The detector arrays are packaged along with their microwave SQUID multiplexing ($\mu$mux) readout \cite{dober2020, jones2023} and optical coupling elements into detector modules, described in \cite{healy2020, mccarrick2021a}.
The modules are read out using SLAC Microresonator RF (SMuRF) electronics \cite{yu2023}.
SO will initially use 49 detector modules: 10 LF, 26 MF, and 13 UHF, totalling approximately 68,000 TESs.
Prior to installation on the telescopes, the detector modules are evaluated in laboratory cryostats to ensure their performance is sufficient to achieve the sensitivity projections in \cite{thesimonsobservatorycollaboration2019b}.
The first MF detector modules were produced in early 2021, and their initial testing results are reported in \cite{mccarrick2021, wang2022b}.
Here we present the testing results of the full set of 26 SO MF detector modules.

\section{Testing Method}\label{sec:method}
The $\mu$mux circuits for each module are first verified cryogencially independently of the detectors, as described in \cite{mccarrick2021a}.
A vector network analyzer (VNA) is used to measure the properties of the microwave resonators, and the readout noise-equivalent current (NEI) of each multiplexed channel is measured through SMuRF.
After coupling the $\mu$mux circuits to the detector wafers, the completed modules are re-installed into the dilution refrigerator cryostat and screened using the equipment and methods described in \cite{wang2022b}.
One-third of each detector module is exposed to a variable-temperature optical load internal to the cryostat, while the remaining two-thirds are masked.
Sets of TES $I$-$V$ curves are taken on the masked detectors while varying the bath temperature over the range 60--200~mK to yield measurements of TES normal resistance $R_N$, superconducting transition temperature $T_c$ and saturation power $P_{b50}$\footnotemark{}.
Further sets of $I$-$V$ curves are taken on all detectors while varying the cold load temperature over the range 8--20~K.
The data from the masked detectors are used as a control, while the absorbed power in the optical detectors is compared to the change in emitted power over the designed detector passband to yield measurements of detector optical efficiency $\eta_{\mathrm{opt}}$.

Measurements of detector noise and effective thermal time constants $\tau_{\mathrm{eff}}$ are made at multiple points in the TES transition.
The value of $\tau_{\mathrm{eff}}$ is extracted from the detector response to a square-wave input on the TES bias line, which also provides a measure of the TES responsivity \cite{niemack2008}.
The TES responsivity and current noise are combined to yield a measurement of bolometer noise-equivalent power (NEP).

\footnotetext{In this work, we define saturation power as the electrical bias power required to bias an optically-dark TES to 50\%~$R_N$ at the intended operating bath temperature of 100~mK. We denote this quantity as $P_{b50}$.}

\section{Results and Discussion}\label{sec:results}

\subsection{Resonator Properties}\label{subsec:res}
\begin{figure}[h]%
\centering
\includegraphics[width=1\textwidth]{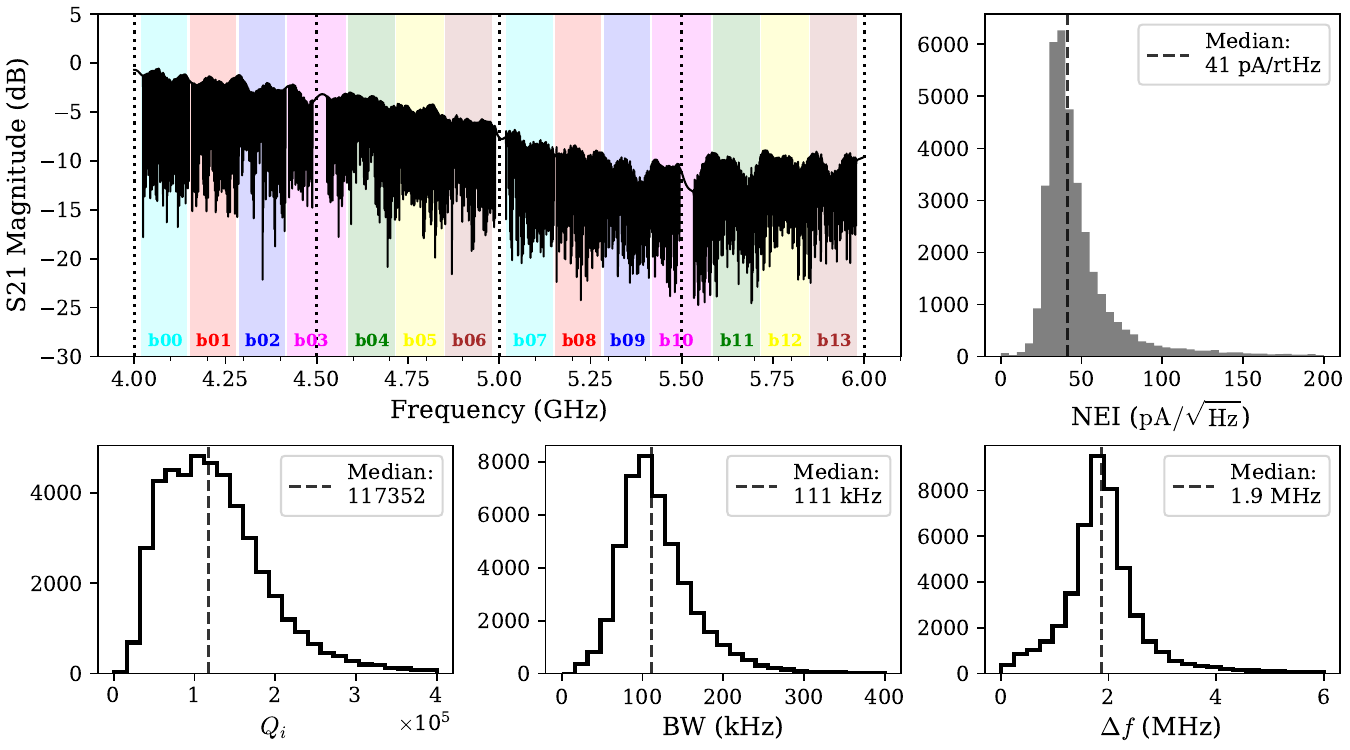}
\caption{\emph{Top left:} Transmission through one readout chain containing 924 resonators, with the nominal frequency ranges of each of the 14 multiplexer chips (b00, b01,...,b13) shown as shaded regions.
Intentional gaps in the frequency schedule are included around the edges of the SMuRF readout bands, marked by the dashed lines.
\emph{Top right:} Histogram of readout-only white-noise levels across all multiplexer channels in the 26 modules prior to connecting TESs.
\emph{Bottom:} Histograms of internal quality factor $Q_i$, bandwidth $BW$, and frequency spacing $\Delta f$, of all resonators across the 26 MF detector modules. Median values are given in the legends.
In these measurements and in operation, the typical tone power at the resonators is -70~dBm.
}\label{fig:res_params}
\end{figure}

An example VNA sweep of an RF readout chain is shown in the top left panel of Figure~\ref{fig:res_params}.
Each readout chain contains 14 daisy-chained $\mu$mux chips, with each chip containing 66 resonators and each detector module utilizing two such readout chains \cite{mccarrick2021a}.
Figure~\ref{fig:res_params} also shows histograms of readout white noise, internal quality factor $Q_i$, bandwidth $BW$, and frequency spacing $\Delta f$ measured across all resonators in all detector modules, with median values indicated.
The average resonator yield per module is 96\%\footnote{Initial module testing often reveals one or more shorted multiplexer chips, which, if no compatible replacement chips are available, are replaced with passive through chips with no resonators. 
If these missing multiplexer chips are omitted from the expected total, the resonator yield increases to 99\%.}.
Of the identified resonators, 90\% have $Q_i>50,000$, an SO criterion to indicate adequate noise performance.
The median $BW$ and  $\Delta f$ values are in good agreement with the target values of 100~kHz and 1.8~MHz, respectively \cite{dober2020}.
For channels read out by SMuRF, the average yield decreases to 89\% due to factors such as frequency collisions, resonators falling outside of the usable readout bandwidth, or poor SQUID response. 
The readout channel white noise distribution has a median of 41~pA/$\sqrt{\mathrm{Hz}}$ and an inverse-variance-weighted average of 35~pA/$\sqrt{\mathrm{Hz}}$.
This is well below the expected on-sky TES-current noise of 160~pA/$\sqrt{\mathrm{Hz}}$, and it achieves the target improvement of NEI $<45$~pA/$\sqrt{\mathrm{Hz}}$ laid out in \cite{mccarrick2021a} to reach goal readout performance requirements.

\subsection{TES Bolometer Properties}\label{subsec:tes}
The detector parameters of each of the SO MF modules are summarized in Table~\ref{tab:detparams}.
The TES yield exceeds the baseline requirement of 70\% on all modules, with an average operable yield of 84\%.
Across the  full distribution, $R_N = 7.5 \pm 0.4$~m$\Omega$, while on individual wafers the standard deviation is $\pm 0.2$~m$\Omega$.
The variation in $R_N$ on each detector wafer is primarily radial, caused by a known variation in layer thickness from the deposition process during detector wafer fabrication.
This radial effect is also present to an even smaller extent in other parameters, such as $T_c$ and $P_{b50}$.

\begin{figure}[h]%
\centering
\includegraphics[width=1\textwidth]{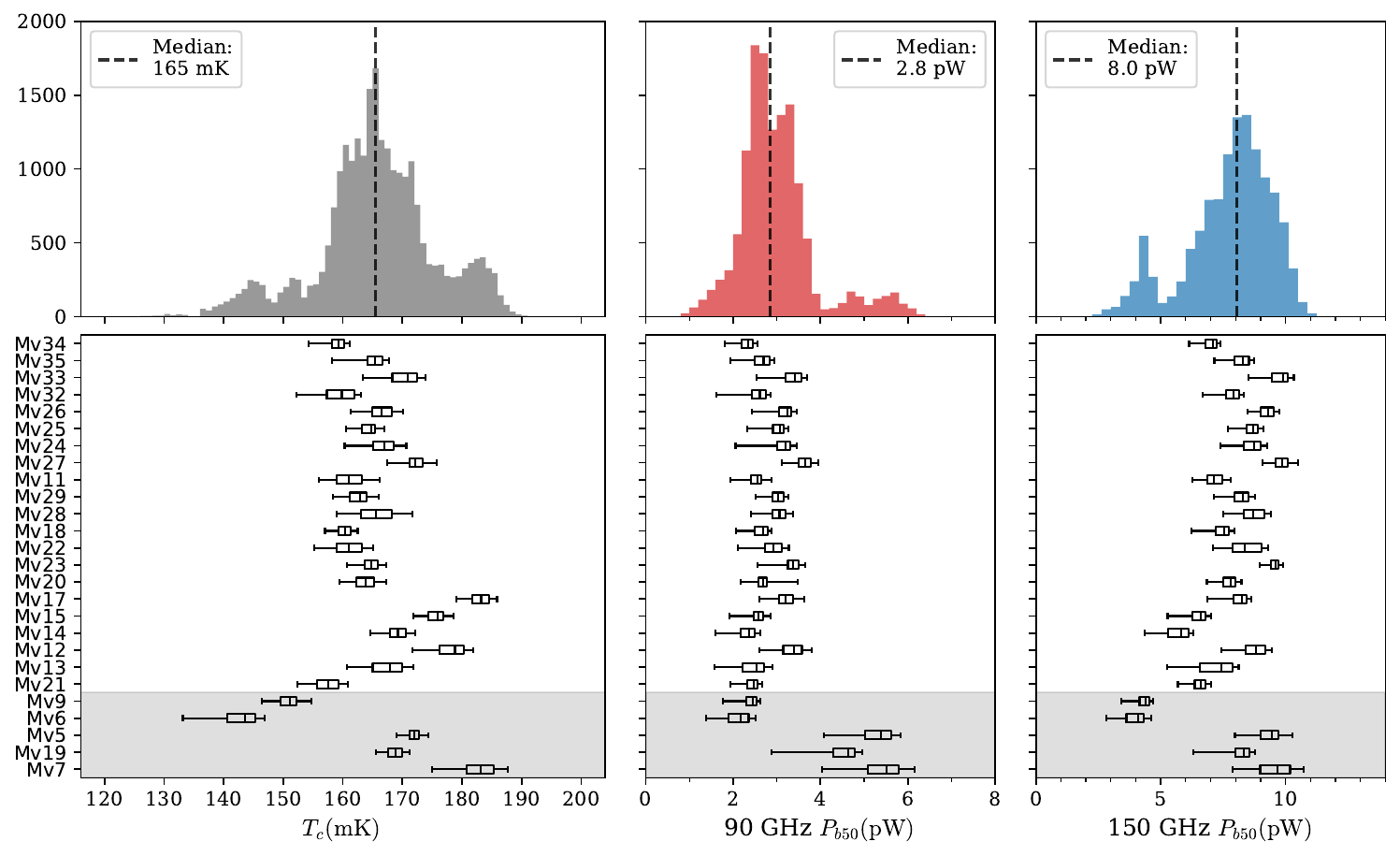}
\caption{Distributions of TES critical temperature $T_c$ and saturation power $P_{b50}$.
The top portion of each plot shows the distribution across all 26 modules.
The lower portions of the plots show the parameter distributions for individual detector modules, where boxes indicate quartiles and whiskers indicate the central 90\% of the data.
The modules are ordered with the newest detector wafers on top and the oldest at bottom; the shaded region indicates the first few wafers made with different targets and design geometries.
The labels at left reflect the internal SO module naming scheme.
}\label{fig:tc-psat}
\end{figure}

The saturation power of the bolometers is targeted to be high enough to prevent the detectors from saturating under expected loading conditions, while not being much higher than this to avoid excess noise.
The TES transition temperature $T_c$ is an important determiner of $P_{b50}$, and the distributions of both of these parameters across all 26 detector modules are plotted in Figure~\ref{fig:tc-psat}.
In total, 73\% of 90~GHz TESs fall within their target range of 2.0--3.3 pW, and 76\% of 150~GHz fall within their target range of 5.4--9.0 pW.
Five of the detector wafers were made with a different design geometry before the target parameters were finalized; these are indicated by the gray shaded region in  Figure~\ref{fig:tc-psat}.
When excluding these prototype wafers, the overall distributions are $T_c = 166 \pm 7$~mK and $P_{b50}=(2.9 \pm 0.6, 8.0 \pm 1.2)$~pW for (90, 150)~GHz, respectively.

\begin{sidewaystable}
\renewcommand\theadfont{\setfsize}
\caption{Detector module parameter summary}\label{tab:detparams}
\begin{tabular*}{\textheight}{@{\extracolsep\fill}lcccccccccccccc}
\toprule%
\thead{Module \&\\Det. Waf. IDs} & \thead{TES\\Yield} & \thead{$R_N$\\(m$\Omega$)} & \thead{$T_\mathrm{c}$ \\ (mK)} & \multicolumn{2}{c}{\thead{$P_{b50}$\\(pW)}} & \multicolumn{2}{c}{$100\eta_{\mathrm{opt}}$} & \multicolumn{2}{c}{\thead{$\tau_{\mathrm{eff}}$\\(ms)}} & \multicolumn{2}{c}{\thead{NEP\\(aW/$\sqrt{\mathrm{Hz}}$)}}
\\\cmidrule{5-6}\cmidrule{7-8}\cmidrule{9-10}\cmidrule{11-12}
& & & & 90 & 150 & 90 & 150 & 90 & 150 & 90 & 150
\\
\midrule
Mv5 (004d-$b$) & 1325 & 7.3 & 172 &5.4 $\pm$ 0.3 & 9.4 $\pm$ 0.3 & 71 $\pm$ 2 & 48 $\pm$ 2 & - & - & - & - \\ 
Mv6 (005e-$c$) & 1575 & 8.0 & 144 &2.2 $\pm$ 0.2 & 4.1 $\pm$ 0.3 & 63 $\pm$ 6 & 52 $\pm$ 4 & - & - & - & - \\ 
Mv7 (001a-$a$) & 1426 & 7.5 & 183 &5.5 $\pm$ 0.3 & 9.7 $\pm$ 0.6 & 93 $\pm$ 6 & 74 $\pm$ 6 & 1.8 $\pm$ 0.2 & 1.7 $\pm$ 0.2 & 21 $\pm$ 6 & 21 $\pm$ 2 \\ 
Mv9 (006f-$c$) & 1572 & 7.8 & 151 &2.4 $\pm$ 0.1 & 4.4 $\pm$ 0.2 & 39 $\pm$ 17 & 46 $\pm$ 8 & 2.0 $\pm$ 0.2 & 1.5 $\pm$ 0.1 & 21 $\pm$ 7 & 20 $\pm$ 3 \\ 
Mv11 (031e-$\theta$) & 1388 & 7.1 & 161 &2.6 $\pm$ 0.1 & 7.1 $\pm$ 0.3 & 81 $\pm$ 2 & 50 $\pm$ 7 & - & - & - & - \\ 
Mv12 (017q-$\delta$) & 1412 & 7.9 & 179 &3.4 $\pm$ 0.2 & 8.8 $\pm$ 0.4 & 85 $\pm$ 3 & 51 $\pm$ 2 & 2.5 $\pm$ 0.8 & 3.4 $\pm$ 0.9 & 15 $\pm$ 3 & 23 $\pm$ 3 \\ 
Mv13 (016p-$\delta$) & 1316 & 7.0 & 168 &2.5 $\pm$ 0.2 & 7.4 $\pm$ 0.5 & 70 $\pm$ 9 & 45 $\pm$ 7 & 2.0 $\pm$ 0.2 & 0.9 $\pm$ 0.1 & 10 $\pm$ 1 & 18 $\pm$ 2 \\ 
Mv14 (018r-$\epsilon$) & 1610 & 8.0 & 169 &2.4 $\pm$ 0.2 & 5.8 $\pm$ 0.3 & 71 $\pm$ 3 & 52 $\pm$ 2 & 2.8 $\pm$ 0.4 & 1.8 $\pm$ 0.2 & 19 $\pm$ 6 & 22 $\pm$ 6 \\ 
Mv15 (019s-$\epsilon$) & 1375 & 7.6 & 176 &2.6 $\pm$ 0.1 & 6.6 $\pm$ 0.2 & 71 $\pm$ 2 & 49 $\pm$ 1 & 2.3 $\pm$ 0.2 & 1.5 $\pm$ 0.1 & 13 $\pm$ 2 & 18 $\pm$ 2 \\ 
Mv17 (020t-$\epsilon$) & 1633 & 7.7 & 183 &3.2 $\pm$ 0.2 & 8.3 $\pm$ 0.2 & 67 $\pm$ 3 & 46 $\pm$ 3 & 2.1 $\pm$ 0.2 & 1.3 $\pm$ 0.1 & 16 $\pm$ 4 & 23 $\pm$ 4 \\ 
Mv18 (030d-$\eta$) & 1463 & 7.1 & 160 &2.7 $\pm$ 0.1 & 7.5 $\pm$ 0.3 & 79 $\pm$ 1 & 54 $\pm$ 2 & 1.8 $\pm$ 0.2 & 0.8 $\pm$ 0.1 & 15 $\pm$ 2 & 24 $\pm$ 2 \\ 
Mv19 (002b-$b$) & 1366 & 7.0 & 169 &4.6 $\pm$ 0.2 & 8.3 $\pm$ 0.3 & 67 $\pm$ 2 & 48 $\pm$ 2 & 1.1 $\pm$ 0.1 & 0.9 $\pm$ 0.1 & 13 $\pm$ 1 & 18 $\pm$ 1 \\ 
Mv20 (024x-$\eta$) & 1655 & 7.4 & 164 &2.7 $\pm$ 0.1 & 7.8 $\pm$ 0.2 & 75 $\pm$ 3 & 48 $\pm$ 6 & 2.3 $\pm$ 0.6 & 0.8 $\pm$ 0.1 & 14 $\pm$ 3 & 21 $\pm$ 2 \\ 
Mv21 (015o-$\delta$) & 1381 & 8.0 & 158 &2.5 $\pm$ 0.1 & 6.6 $\pm$ 0.2 & 81 $\pm$ 2 & 56 $\pm$ 3 & - & - & - & - \\ 
Mv22 (026z-$\eta$) & 1507 & 7.5 & 161 &2.9 $\pm$ 0.2 & 8.4 $\pm$ 0.6 & 78 $\pm$ 3 & 52 $\pm$ 8 & 2.4 $\pm$ 0.5 & 0.8 $\pm$ 0.1 & 15 $\pm$ 2 & 22 $\pm$ 3 \\ 
Mv23 (025y-$\eta$) & 1627 & 7.7 & 165 &3.4 $\pm$ 0.1 & 9.6 $\pm$ 0.1 & 88 $\pm$ 2 & 53 $\pm$ 4 & 2.6 $\pm$ 0.5 & 0.8 $\pm$ 0.0 & 17 $\pm$ 2 & 23 $\pm$ 2 \\ 
Mv24 (033g-$\iota$) & 1497 & 7.9 & 167 &3.2 $\pm$ 0.2 & 8.7 $\pm$ 0.3 & 82 $\pm$ 2 & 57 $\pm$ 1 & 1.8 $\pm$ 0.3 & 0.8 $\pm$ 0.1 & 15 $\pm$ 2 & 28 $\pm$ 3 \\ 
Mv25 (034h-$\iota$) & 1582 & 7.8 & 165 &3.1 $\pm$ 0.1 & 8.7 $\pm$ 0.2 & 77 $\pm$ 1 & 59 $\pm$ 2 & 2.1 $\pm$ 0.3 & 1.2 $\pm$ 0.1 & 15 $\pm$ 2 & 27 $\pm$ 3 \\ 
Mv26 (035i-$\iota$) & 1567 & 7.8 & 167 &3.2 $\pm$ 0.1 & 9.3 $\pm$ 0.3 & 89 $\pm$ 3 & 60 $\pm$ 2 & 1.5 $\pm$ 0.2 & 0.8 $\pm$ 0.0 & 17 $\pm$ 3 & 29 $\pm$ 3 \\ 
Mv27 (032f-$\iota$) & 1611 & 8.2 & 172 &3.7 $\pm$ 0.1 & 9.9 $\pm$ 0.2 & 84 $\pm$ 2 & 49 $\pm$ 4 & 2.2 $\pm$ 0.3 & 1.2 $\pm$ 0.2 & 13 $\pm$ 2 & 21 $\pm$ 2 \\ 
Mv28 (027a-$\theta$) & 1442 & 7.2 & 166 &3.1 $\pm$ 0.2 & 8.7 $\pm$ 0.4 & 88 $\pm$ 3 & 58 $\pm$ 4 & 1.9 $\pm$ 0.2 & 0.9 $\pm$ 0.1 & 12 $\pm$ 2 & 21 $\pm$ 2 \\ 
Mv29 (028b-$\theta$) & 1353 & 7.2 & 163 &3.0 $\pm$ 0.1 & 8.3 $\pm$ 0.3 & 87 $\pm$ 2 & 60 $\pm$ 5 & 1.7 $\pm$ 0.2 & 0.9 $\pm$ 0.1 & 13 $\pm$ 2 & 22 $\pm$ 2 \\ 
Mv32 (036j-$\kappa$) & 1539 & 7.3 & 160 &2.6 $\pm$ 0.2 & 7.9 $\pm$ 0.3 & 86 $\pm$ 3 & 62 $\pm$ 1 & 1.8 $\pm$ 0.2 & 0.8 $\pm$ 0.1 & 17 $\pm$ 4 & 23 $\pm$ 4 \\ 
Mv33 (037k-$\kappa$) & 1535 & 7.3 & 171 &3.4 $\pm$ 0.2 & 9.9 $\pm$ 0.3 & 87 $\pm$ 2 & 57 $\pm$ 2 & 1.4 $\pm$ 0.2 & 0.8 $\pm$ 0.0 & 13 $\pm$ 1 & 23 $\pm$ 2 \\ 
Mv34 (039m-$\lambda$) & 1277 & 7.0 & 159 &2.3 $\pm$ 0.1 & 7.1 $\pm$ 0.2 & 79 $\pm$ 3 & 65 $\pm$ 3 & 2.0 $\pm$ 0.2 & 0.9 $\pm$ 0.1 & 13 $\pm$ 2 & 24 $\pm$ 6 \\ 
Mv35 (038l-$\lambda$) & 1465 & 7.2 & 165 &2.7 $\pm$ 0.2 & 8.3 $\pm$ 0.3 & 77 $\pm$ 3 & 62 $\pm$ 1 & 2.0 $\pm$ 0.2 & 0.8 $\pm$ 0.1 & 14 $\pm$ 2 & 24 $\pm$ 3 \\
\botrule
\end{tabular*}
\footnotetext{Note: TES yield is the number of operable readout channels that exhibited transitions in their $I$-$V$ curves during testing.
Other entries in the table are given as the median $\pm$ the median absolute deviation, split by observing band where appropriate.
The variations in $R_N$ and $T_c$ are all approximately  $\pm$ 0.2~m$\Omega$ and $\pm$ 3~mK, respectively, and are therefore suppressed in the table.
Saturation power $P_{b50}$, effective thermal time constant ($\tau_{\mathrm{eff}}$) and noise-equivalent power (NEP) are measured in transition at 40\%--60\% $R_N$ with no incident optical power.
Not all data are available for all detector modules.}
\end{sidewaystable}

\begin{figure}[h]%
\centering
\includegraphics[width=1\textwidth]{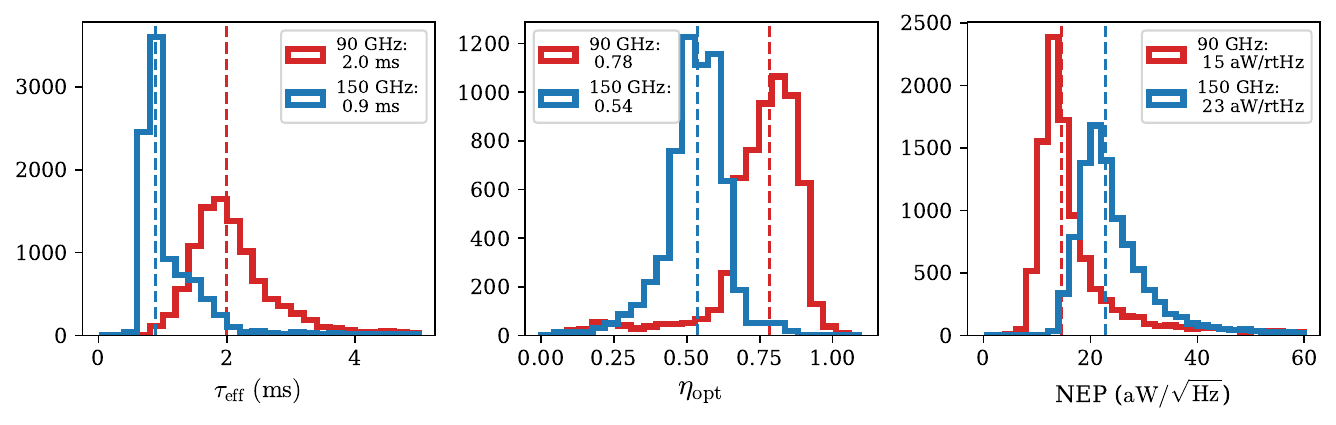}
\caption{Histograms of detector time constant $\tau_{\mathrm{eff}}$, optical efficiency $\eta_{\mathrm{opt}}$, and noise-equivalent power (NEP) across all detector modules. $\tau_{\mathrm{eff}}$ and NEP are measured in transition at 40\%--60\% $R_N$ with no optical loading. Median values are given in the legends.}\label{fig:param_hists}
\end{figure}

Histograms of $\tau_{\mathrm{eff}}$, $\eta_{\mathrm{opt}}$, and NEP are shown in Figure~\ref{fig:param_hists}.
The NEP and $\tau_{\mathrm{eff}}$ values are measured with the detectors biased in the range 40\%--60\%~$R_N$ with the cold stage at 100~mK and no optical loading.
The main contributions to NEP in this configuration are from readout noise and thermal noise, and in the baseline model used in \cite{thesimonsobservatorycollaboration2019b}, these have an expected quadrature sum of approximately (13, 22)~aW/$\sqrt{\mathrm{Hz}}$  at (90, 150) GHz.
While the peaks of the measured NEP distributions align with this, the median values are dragged high by the long tails.
However, if the full distribution is used to compute an effective per-detector NEP, and this value compared to the model\footnotemark{} used in \cite{thesimonsobservatorycollaboration2019b}, we still expect to achieve baseline sensitivity targets with some margin.
The detector $\eta_{\mathrm{opt}}$ values are expected to be in the range of 0.7--0.8, while their measured distributions have medians (interquartile ranges) of 0.78 (0.70--0.85) for 90~GHz and 0.54 (0.48--0.59) for 150~GHz.
The cause of the tail of $\eta_{\mathrm{opt}}$ for 90~GHz, and the overall lower $\eta_{\mathrm{opt}}$ for 150~GHz, is unknown.
However, projections accounting for the lower values still indicate MF performance will exceed the baseline projections in \cite{thesimonsobservatorycollaboration2019b}.
The $\tau_{\mathrm{eff}}$ distributions have medians (interquartile ranges) of 2.0~ms (1.7--2.4)~ms for 90~GHz and 0.9~ms (0.8--1.2)~ms for 150~GHz, whereas the target maximum for both frequencies is $\tau_{\mathrm{eff}}\sim$1~ms, set by a combination of beam size and telescope scanning speed as well as polarization sensitivity considerations \cite{simon2014}.
This disagreement is being addressed in future fabrication batches by reducing the heat capacity of the TES island and increasing the 90~GHz saturation powers to be closer to those of the prototype MF wafers.
These wafers may be used as replacements in the future if required.

\footnotetext{The SO MF baseline model assumed a larger pixel size, and therefore fewer detectors per wafer, than the as-built version. The realized percentage detector yield (84\%) also exceeds that used in the model (70\%). The larger detector count offsets the increase in NEP.
}

\section{Conclusion}\label{sec:conclusion}

We have presented the laboratory characterization of the 26 MF detector modules to be deployed in the first phase of SO, using measurements from over 38,000 TESs.
Overall, the data show high yield and uniformity on the detector wafers and impressive process control across fabrication batches at high levels of production.
The performance of the detector modules are consistent with the instrument sensitivity projections in \cite{thesimonsobservatorycollaboration2019b}.
The SO MF detector modules are being deployed to the field now, with initial commissioning observations underway and full science observations with SO expected to begin in 2024.
Looking beyond the first phase of SO, the Advanced Simons Observatory upgrade (NSF Award \#2153201) will build and deploy 18 additional detector modules with 30,000 more TESs, doubling the mapping speed of CMB temperature and polarization anisotropies at high resolution.

\backmatter

\bmhead{Acknowledgments}
This work was supported in part by the Simons Foundation (Award \#457687, B.K.).

\bibliography{LTD20}

\end{document}